\documentclass[b5paper,twoside]{jpconf}  
\usepackage{graphicx}
\usepackage{pstricks}
 \usepackage[small]{caption}

\newcommand{\mstar}{M$_\odot$}
\newcommand{\spitzer}{{\it Spitzer}}
\newcommand{\microm}{$\mu$m}
\newcommand{\disperse}{DisPerSE}
\newcommand{\herschel}{{\it Herschel}}
\newcommand{\Av}{{$A_{\mathrm{V}}$}}
\newcommand{\arcsec}{{$^{''}$}}

\begin{document}

\title[Recent results from HOBYS]{HOBYS' view of Vela~C and W48: a ridge and a mini-starburst}  

\author[T. Hill et~al., ]{T. Hill$^1$, Q. Nguy$\tilde{\hat{\rm e}}$n Lu{\hskip-0.65mm\small'{}\hskip-0.5mm}o{\hskip-0.65mm\small'{}\hskip-0.5mm}ng$^1$, F. Motte$^1$, P. Didelon$^1$, V.~Minier$^1$ and the HOBYS consortium}

\address{$^1$ Laboratoire AIM, CEA/IRFU CNRS/INSU Universit\'e Paris Diderot, CEA-Saclay, 91191 Gif-sur-Yvette Cedex, France}

\ead{tracey.hill@cea.fr} 

\begin{abstract}
We present recent results from the \herschel\ HOBYS guaranteed time key program of the Vela~C and W48 star-forming complexes. 
 We examine the column density distribution in Vela~C, in particular focusing on the cloud structure using probability distribution functions, and characterise the star formation efficiency in W48.
\end{abstract}

\vspace{-0.5cm}
\section{HOBYS \& star-forming complexes }

The \herschel\ imaging survey of OB young stellar objects (HOBYS) 
capitalises on the unprecedented mapping capabilities of the \herschel\ space observatory \cite{griffin10} to compile the first complete, systematic and unbiased sample of nearby ($<$\,3\,kpc) high-mass star progenitors \cite{motte10, hill10b}. 
The Vela~C giant molecular cloud complex, is an ideal laboratory in which to study star formation. Vela~C is particularly unusual owing to its proximity (700\,pc) {\it and} the fact that it houses
 high, intermediate and low mass star formation \cite{massi03} and so may provide clues as to what causes the different modes of star formation. The IRDC G035.39--00.33 is an infrared dark cloud filament located in W48 (3 kpc), with a mass and median column density among the highest of known infrared dark clouds \cite{peretto10}.

\section{Observations and Column Density maps}

Vela~C and W48 
were observed on 2010, May 18 and September 18--19, respectively, using the parallel PACS/SPIRE mode at 70/160\microm\ and 250/350/500\microm\ and a scan speed of 20\arcsec/s. The data reduction 
is as described in \cite{hill11} and \cite{quang11}, respectively.
The column density and dust temperature maps were drawn from fitting pixel-by-pixel spectral energy distribution (SEDs),
see \cite{hill11}.

\section{Vela~C}

The column density and temperature maps reveal the southern part of Vela~C to be dominated by cold, dense material, while the centre of the map has a cold dense filament ($\sim$15--18\,K) coexisting with warmer and less dense material.

\begin{figure}[h]
\begin{center}
\hspace{1.1cm}\includegraphics[height=4.1cm]{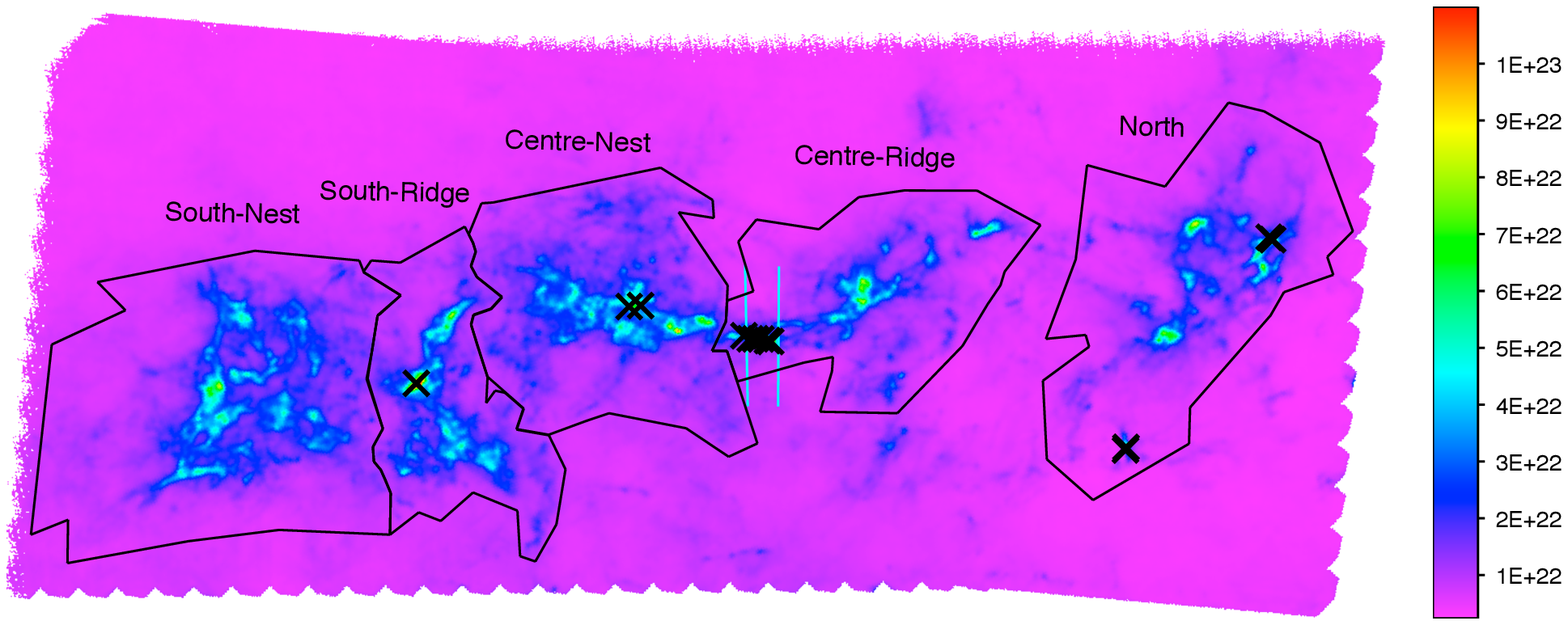}
\includegraphics[angle=270,width=9cm]{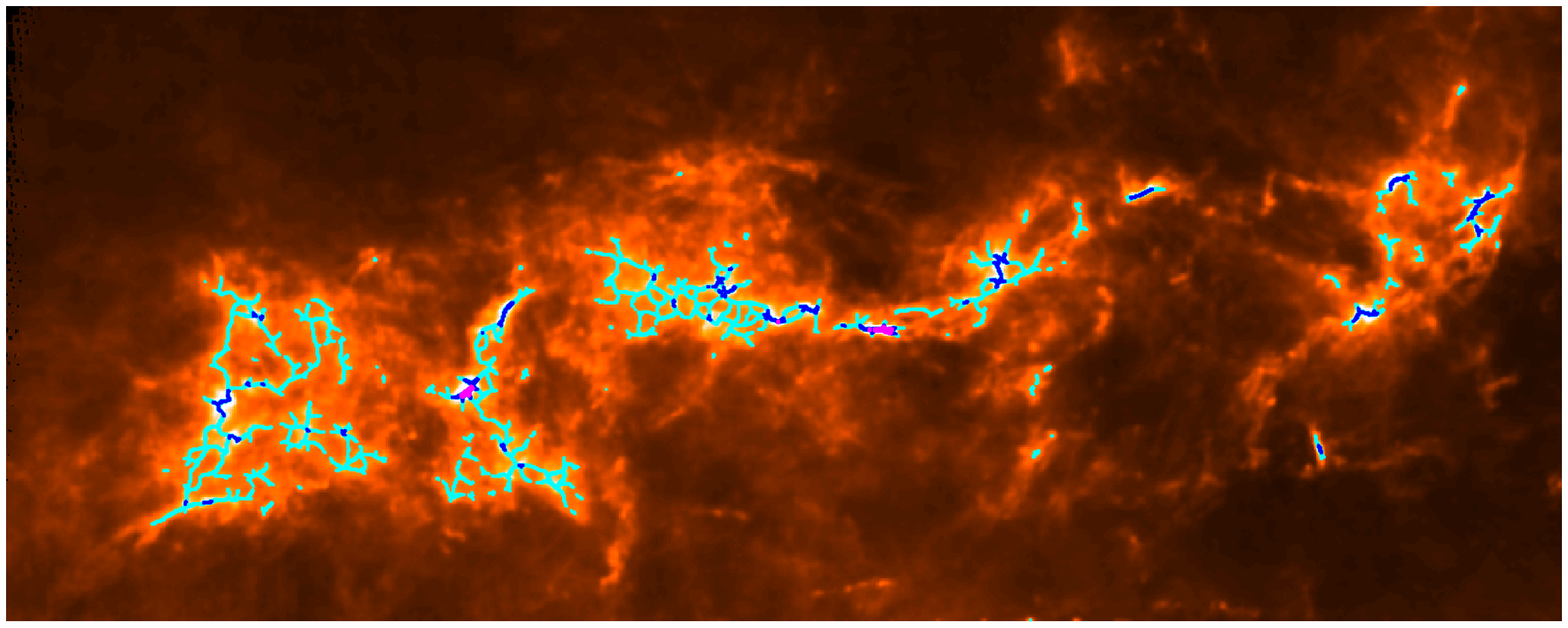}
\begin{pspicture}(0,0)
\psline[linecolor=cyan, linewidth=1.5pt](-9,-0.2)(-8.5,-0.2)
\rput[linecolor=cyan](-8.2,-0.2){\white \tiny \Av\ 25 mag}
\psline[linecolor=blue, linewidth=1.5pt](-9,-0.4)(-8.5, -0.4)
\rput[linecolor=blue](-8.2,-0.4){\white \tiny \Av\ 50 mag}
\psline[linecolor=magenta,linewidth=1.5pt](-9,-0.6)(-8.5,-0.6)
\rput[linecolor=magenta](-8.2,-0.6){\white \tiny \Av\ 100 mag}
\end{pspicture}
\caption{\footnotesize Top: Column Density map with sub-regions, as defined at an \Av $>$ 7 mag, overlaid. The black crosses are the 13 most massive sources with (S:N$>$ 50, and Mass 20--60\mstar).
Bottom: Column density with the filamentary structure detected by \disperse. Ridges appear magenta.  
 \label{fig:vela:dens}}
\vspace{-0.5cm}
\end{center}
\end{figure}

At an \Av\ magnitude of 7, the Vela~C complex segregates into five distinct sub-regions (Fig. \ref{fig:vela:dens}) containing $\sim$\,4.2\,$\times$10\,$^6$\,\mstar, each with different
characteristics and filamentary structure. Recent results of the Gould Belt programme 
revealed the presence of star-forming supercritical (gravitationally unstable) filamentary structures above an \Av\ $>$ 7 mag \cite{andre11}. 
To trace the filaments evident in Vela~C the discrete persistent structure
extractor (\disperse) was applied to the map \cite{sousbie11} - Fig. \ref{fig:vela:dens}. The filamentary networks in Vela~C only differentiate at high \Av. Only two notable filamentary structures reach a high column density of 10$^{23}$\,cm$^{-2}$, which we call `ridge'.
We stress that a ridge is not simply a peak of column density which
is attributed to a single core/clump, but rather a filament of a characteristic width.
The central ridge in Vela~C has an average column density $\sim$\,10$^{22.5}$\,cm$^{-2}$ over a width of $\sim$0.3\,pc and length $\sim$\,1.5\,pc. 

Thirteen high-mass compact sources were identified in Vela~C, their mass (20\,--\,60\,\mstar) and size (0.03 -- 0.2\,pc) qualifies them as starless or protostellar dense cores. The majority of 
them are found in ridges, see \cite{hill11}, making ridges the preferential sites of high-mass star formation, at least in Vela~C.

\begin{figure}
\begin{center}
\includegraphics[height=5.5cm]{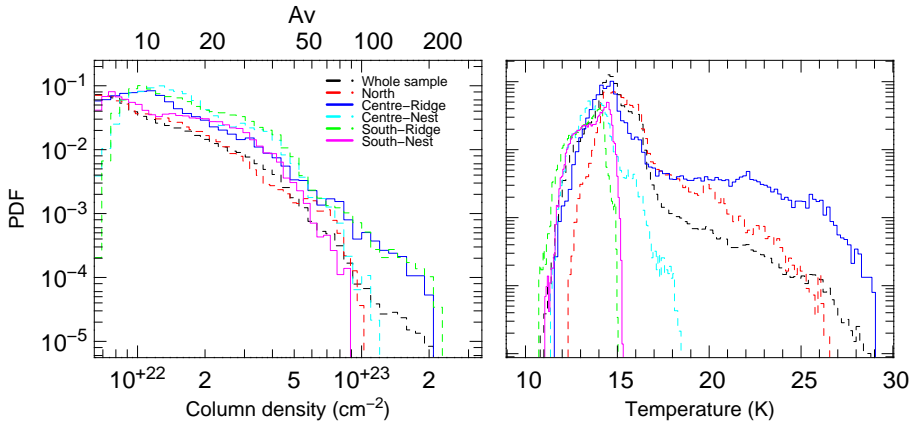}
\caption{
\footnotesize Normalised column density (left) and temperature (right) probability distribution functions (PDFs) of Vela C and its sub-regions.
Note that the Centre-Ridge has a flatter column density slope than the
South-Nest and a bimodal temperature distribution. \label{fig:pdf} }
\end{center}
\vspace{-0.5cm}
\end{figure}

Column density probability distribution functions (PDFs) have been
 used to characterise cloud structure as a function of star formation activity. 
\cite{kain09} find that the shape of their observed column density 
PDF reflects the star formation status, 
 with active star 
forming regions having a high column density power-law tail.
The {\it striking} difference between the flat tail of the Centre-Ridge and the steepening of the South-Nest (Fig. \ref{fig:pdf}) reflects the presence 
of a ridge. 
In some models, tails are indicative of gravity-dominated regions \cite{klessen00}. Above \Av\ $>$ 30 mag, the Centre-Ridge reaches a higher column density with a flatter PDF than the South-Nest. A flatter slope is expected for coherent structures created via constructive large-scale flows rather than small scale turbulence \cite{fed10}. A higher column density results when the compression of material is stronger and thus the velocity difference of converging flows is probably higher.  The steeper PDF and the smaller concentration of material in the South-Nest \cite{hill11}, with respect to the Centre-Ridge, suggests this region to be more turbulent.

\section{IRDC G035.39-00.33 in the W48 complex}

The IRDC G035.39--00.33 is a prominent elongated structure oriented north-south in the \herschel\ map of W48. It appears as a dark feature shortward of 70\microm, and in emission at wavelengths longer than 160\microm. The IRDC, $\sim$\,6\,pc in length and $\sim$\,1.7\,pc in width, houses dense (N$_{H_2}$ $\sim$\,3--9\,$\times$ 10$^{22}$ cm$^{-2}$) and cold material (T$_{\rm dust}$ $\sim$\,13\,--\,16\,K). Based on the column density map, this IRDC has a total mass of $\sim$\,5000\,\mstar\ within an area of $\sim$\,8\,pc$^2$.

Fifteen (55\%) dense cores ($<$\,20\,\mstar) and nine (70\%) massive dense cores (MDCs; $<$\,20\,\mstar) of those found in the W48 complex, are located inside the IRDC filament detected by \herschel. With a mass density of 600\,\mstar~pc$^{-2}$ and mass per unit length of $\sim$\,800\,\mstar~pc$^{-1}$, G35.39-00.33 has a much higher density and mass per unit length than equivalent sized regions in Gould Belt clouds. 
 Such a high concentration of MDCs gives a star formation rate (SFR) of 300\,\mstar/yr and a SFR density of 40\,\mstar~yr$^{-1}$\,kpc$^{-2}$ and indicates that the gravitational potential of G035.39--00.33 helped them to build up. Spectral energy distributions indicate these MDCs to be IR-quiet protostellar dense cores harbouring young protostars \cite{quang11}. The abundance of IR-quiet MDCs in G035.39-00.33 suggests that they were formed simultaneously and likely immediately following the formation of the filament. Based on the mass of the filament, 5000\,\mstar, and assuming a standard initial mass function, the star formation efficiency (SFE) of this filament could be as high as 15\%. To reconcile this high SFE with that seen in other complexes, the star formation event would need to be short $\sim$\,10$^6$\,yrs. Such a mini star burst is consistent with the filament being formed through a rapid process, for example by converging flows.

Widespread SiO emission ($\sim$\,0.01--0.05\,Kkms$^{-1}$) was found toward this filament \cite{js10}. Two of the three SiO peaks coincide with IR-quiet MDCs: \#6 (coincident with a water maser) and \#17 (likely hosts an early-stage protostar). In these instances, the SiO emission could easily originate from shocks within protostellar outflows.
 The final (eastern) peak does not correlate with any reliable protostar candidate. A further extraction at this SiO peak was performed and analysis of both \herschel\ and \spitzer\ data indicate that this `source' could be at most a 1\,\mstar\ dense core containing an evolved low-mass class I protostar. Such a protostar would likely not drive a strong enough outflow to cause the  SiO emission.
Since no definite protostar is detected by \herschel\ toward the Eastern SiO peak, the interpretation that the SiO emission could partly originate from large-scale converging-flows may be valid, at least at this location.

\begin{figure}
\begin{center}
\includegraphics[width=13.2cm]{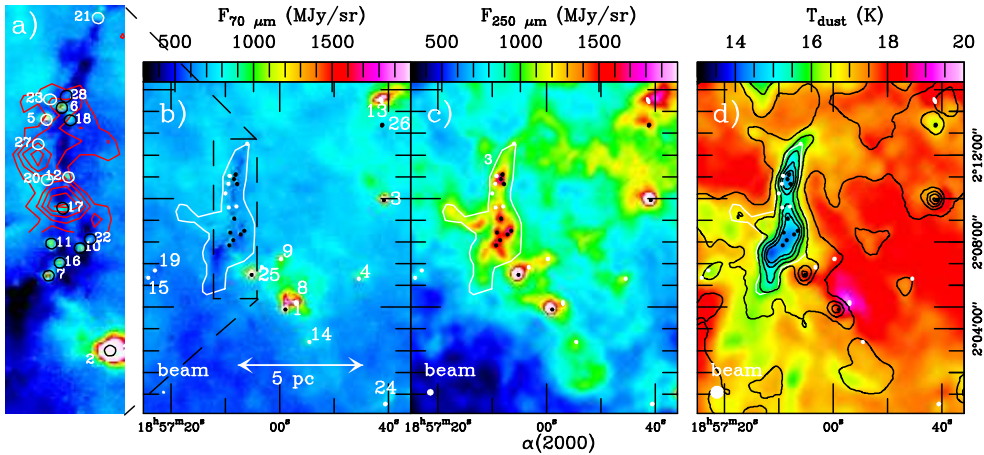}
\vspace{-0.8cm}
\caption{\footnotesize IRDC G035.39-00.33 a) 70\microm\ with SiO contours. b) 70\microm\ image. c) 250\microm\ image d) temperature map with column density contours (1.5\,--\,9 by $1.5~\times~10^{22}$\,cm$^{-2}$). MDCs %
are indicated in black and dense cores
in white. The white polygon is the extent of the IRDC.
}

\vspace{-0.5cm}
\end{center}
\end{figure}

\section{Conclusion}

The HOBYS programme offers valuable insight into nearby star forming complexes. In Vela~C, high-mass star formation proceeds preferentially in high-column density ridges which may result from the constructive convergence of flows. In W48, the IRDC G035.39-00.33 is likely undergoing a mini star-burst of star formation, while the SiO is likely originating from two IR-quiet MDCs, though part of it may originate from low-velocity shocks within converging flows.

\end{document}